\documentclass[a4paper]{article}

\usepackage{INTERSPEECH2022}
\usepackage[symbol]{footmisc}
\usepackage{multirow}
\usepackage{cite}

\title{Generalizing RNN-Transducer to Out-Domain Audio via Sparse Self-Attention Layers}
\name{Juntae Kim$^{1}$\textsuperscript{*}, Jeehye Lee$^{2}$\textsuperscript{*}}

\address{
  $^1$SK Telecom, Seoul, Republic of Korea\\
  $^2$Kakao Enterprise Corporation, Seongnam, Republic of Korea}
\email{jtkim@kaist.ac.kr, jessie.2@kakaoenterprise.com}

\begin{document}

\maketitle
\let\thefootnote\relax\footnotetext{* : These authors have contributed equally to this work}

\begin{abstract}

Recurrent neural network transducer (RNN-T) is an end-to-end speech recognition framework converting input acoustic frames into a character sequence. The state-of-the-art encoder network for RNN-T is the Conformer, which can effectively model the local-global context information via its convolution and self-attention layers. Although Conformer RNN-T has shown outstanding performance, most studies have been verified in the setting where the train and test data are drawn from the same domain. The domain mismatch problem for Conformer RNN-T has not been intensively investigated yet, which is an important issue for the product-level speech recognition system. In this study, we identified that fully connected self-attention layers in the Conformer caused high deletion errors, specifically in the long-form out-domain utterances. To address this problem, we introduce sparse self-attention layers for Conformer-based encoder networks, which can exploit local and generalized global information by pruning most of the in-domain fitted global connections. Also, we propose a state reset method for the generalization of the prediction network to cope with long-form utterances. Applying proposed methods to an out-domain test, we obtained 27.6\% relative character error rate (CER) reduction compared to the fully connected self-attention layer-based Conformers.

\end{abstract}

\section{Introduction}

The recent research focus of automatic speech recognition (ASR) is end-to-end (E2E) frameworks \cite{Chan-LAS16,Chorowski-ABM15,Watanabe-HCA17,Chiu-SSR18,Chiu-MCA18,Graves-CTC06,Graves-TEE14,Graves-SRD13,Rao-EAD17,Sainath-ASO20,Kim-ART20,Daniel-SA19,Christoph-RWTH19,Moritz-CMR21}, which can directly map incoming speech signals into characters \cite{Amodei-DS216} or word targets \cite{Hagen-NSR17,Audhkhasi-DAW17}. The E2E frameworks include encoder--decoder networks \cite{Chan-LAS16,Chorowski-ABM15,Watanabe-HCA17,Chiu-SSR18,Chiu-MCA18}, connectionist temporal classification (CTC) \cite{Graves-CTC06,Graves-TEE14}, and recurrent neural network transducer (RNN-T) \cite{Graves-SRD13,Rao-EAD17,Sainath-ASO20,Kim-ART20}. Considering the various E2E frameworks, RNN-T based approaches have shown promising results based on word error rate (WER) and decoding speed. Most of studies \cite{Anmol-Conformer20,Xiong-ECP21,shim-URS22,Burchi-EC21} on RNN-T have been conducted in the same domain, such as Librispeech \cite{Panayotov-Librispeech15}; however, the robust performance across the different domains must be involved for the production-level ASR. This indicates that it is important to address the domain-mismatch problem between training and inference.

In \cite{Chiu-RMF21}, the domain mismatch problem for RNN-T was intensively investigated. First, the encoder network in the RNN-T suffers from overfitting to the training domain, referred to as the in-domain. Second, the RNN-T is vulnerable to long-form utterances during inference because it is generally trained on short segments. These two problems cause high deletion errors when decoding is conducted on out-domain long-form utterances. To overcome these problems, multiple regularization methods (e.g., variational weight noise \cite{Graves-PVI11}) and dynamic overlapping inference (DOI) which splits long-form utterances into several overlapping segments, have been proposed. Although applying the methods \cite{Chiu-RMF21} improved the WER for out-domain long-form utterances, the investigation was conducted only on long short-term memory (LSTM)-based encoder networks. Furthermore, the long segment length for DOI ($>20$ s) was not investigated although the LSTM was designed for long-term context information \cite{Hochreiter-LSTM97}.

In \cite{Li-ABF21,Xiong-ECP21}, a Conformer-based encoder network was proposed for the RNN-T as the Conformer can effectively model the local-global context information through its convolution and self-attention layers \cite{Anmol-Conformer20}, showing promising performance. However, the domain mismatch problem for Conformer-based encoder network has not been intensively investigated yet.

%was restricted to only the use of local self-attention instead of full context self-attention to improve the generalization ability to cope with long-form utterances.

In this study, we propose a generalization strategy for RNN-T with a Conformer-based encoder network. The main contributions of this study are as follows. (i) The sparse self-attention layers that can exploit both local and global connections are designed for the Conformer: the generalized global connections robust to the domain mismatch problem are identified by pruning most of the redundant global connections while conserving the important global connections considered by the model. 
(ii) The state reset method is proposed to cope with long-form utterances by re-initializing the LSTM states of the prediction network when silence is detected during the decoding phase. Considering the experimental evaluations, we found that combining local and sparse global connections outperformed the local connections alone, and the state reset method showed further improvement in the out-domain test set.

%(ii) The state reset method for the prediction network is proposed to cope with long-form utterances by re-initializing the LSTM states of the prediction network when silence is detected during the decoding phase. Considering the experimental evaluations, we found that combining local and sparse global connections outperformed the local connections alone, and the state reset method showed further improvement in the out-domain test set.

\section{Proposed Methods}

The RNN-T consists of an encoder, prediction, and joint-networks, transcribing acoustic frames into output tokens (e.g., a word-piece unit). The details of the RNN-T can be found in \cite{Graves-SRD13}. In this study, a Conformer, LSTM, and feed-forward network are used for the encoder, prediction, and joint networks.
%The Conformer is a state-of-the-art architecture for RNN-T's encoder network because it has a powerful local-global context modeling ability via its self-attention and convolution layers \cite{Anmol-Conformer20}.

%It transcribes $\textbf{x}=\left\{x_t\right\}_{t=1}^T$ into $\textbf{y}=\left\{y_u\right\}_{u=1}^U$, where $x_t$ is an acoustic frame and $y_u$ is an output token (e.g., a word-piece unit). To deal with the length difference problem between \textbf{x} and \textbf{y}, RNN-T adopts the blank token $\phi$, which enables the RNN-T to decide whether to produce the output token during the \textit{T}-step decoding procedure. Further details of the RNN-T can be found in [8]. In this study, a Conformer, LSTM, and feed-forward network are used for the encoder, prediction, and joint networks, respectively. The Conformer is a state-of-the-art architecture for RNN-T's encoder network because it has a powerful local-global context modeling ability via its self-attention and convolution layers [22].

\begin{figure}[t]
  \centering
  \includegraphics[width=7.2cm]{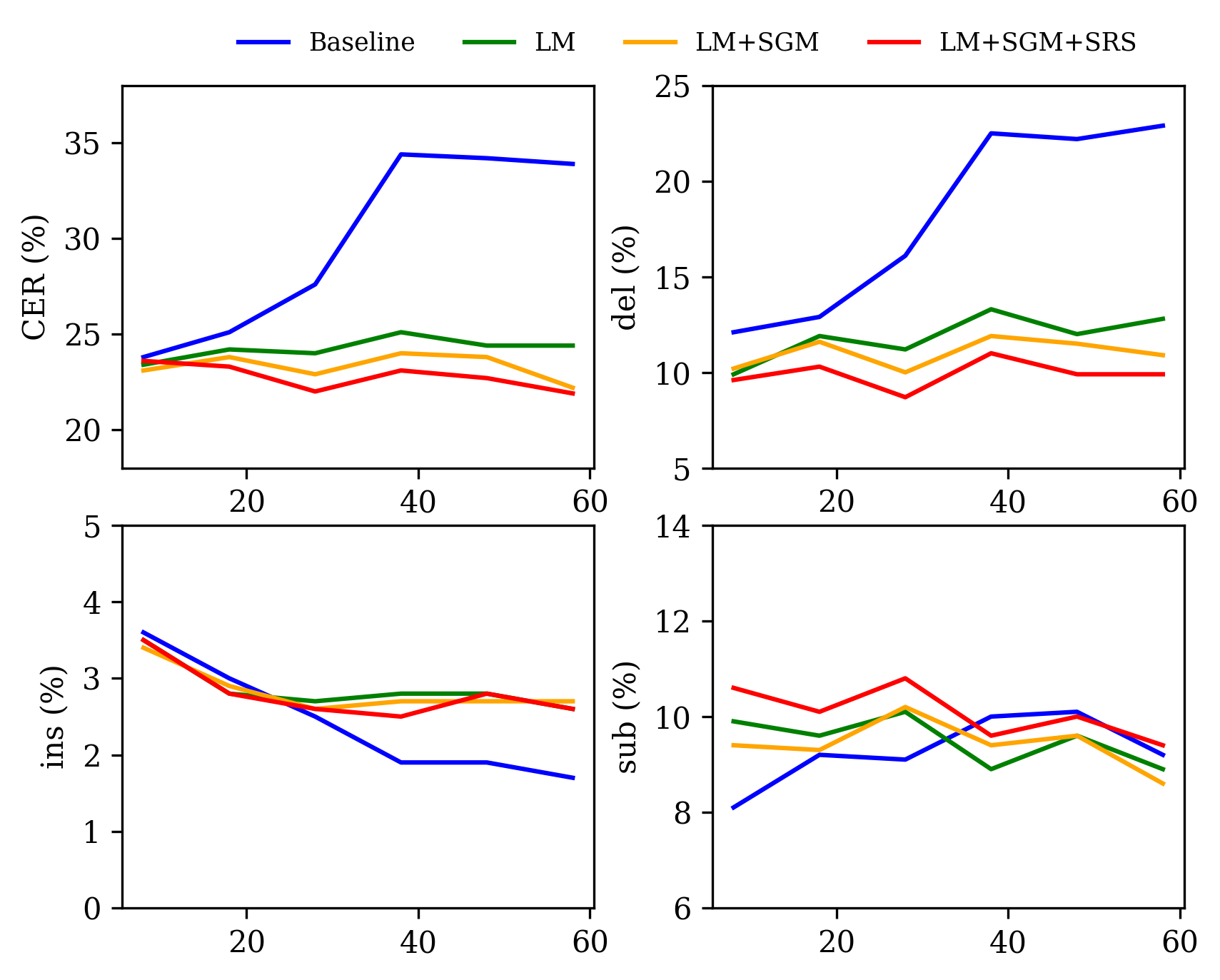}
  \caption{CERs and the deletion, insertion, and substitution errors for Conformer RNN-T (Baseline), with local attention masks (LM); LM and sparse global masks (LM+SGM); LM+SGM and a state reset method at the silence (LM+SGM+SRS) per DOI-8, -18, -28, -38, -48, and -58.}
  \label{figure1}
\end{figure}

\subsection{Sparse self-attention layer for encoder}
\subsubsection{Motivation}
%The vanilla self-attention in the Conformer has full connectivity between queries and keys; most of them are redundant [23--25]. Thus, pruning some connections (i.e., injecting sparsity into the self-attention layer) can improve the generalization ability of the Conformer-based encoder networks. 

Improving the encoder network’s generalization ability is crucial as the RNN-T’s encoder network can easily overfit the in-domain data compared to the other RNN-T components (prediction and joint networks). This leads to excessive deletion errors in the out-domain data, specifically for long-form utterances \cite{Chiu-RMF21}. In \cite{Chiu-RMF21}, they used DOI to split long-form utterances into short-form to improve generalization ability. 

Motivated by the results in \cite{Chiu-RMF21}, we investigated our Conformer RNN-T using the fraction of out-domain test set per the DOI length to deal with the long-form utterance as shown in Figure~\ref{figure1}. Note that all experiments in this work are conducted on Korean corpus, we adopt character error rate (CER) for the evaluation metric. Also, DOI length includes segment and overlap length, e.g., 20 s DOI length (DOI-20) implies 16 s segment and 2 s overlap. Note that we used 2 s overlap for DOI for all experiments in this work. In Figure~\ref{figure1}, although we used the DOI, we found that the deletion error of Conformer RNN-T drastically increased when the DOI length is longer than 20 s. As the deletion error was low with a short DOI length, we assume that the local attention within the Conformer is more generalized than the global attention; thus, we applied the local attention masks (LM) to the self-attention layers in the Conformer. This makes the Conformer only use 2.4 s of past and future context information, which was found from our validation set. As expected, LM applied Conformer RNN-T in Figure~\ref{figure1} showed a stable deletion error per the DOI length.

%Note that the best CER of LM applied RNN-T with Conformer was shown at 60 s DOI length because the prediction network could utilize the fluent long-term context information while the Conformer was restricted by LM. 

%they used only local self-attention to generalize the attention connectivity to deal with long-form utterances. This is an appropriate sparsity pattern as a speech frame is highly correlated with its adjacent frames.

However, one of the strengths of the self-attention layer is that it can jointly learn both local and global patterns from the input features \cite{Vaswani-AAN17}. Furthermore, the Conformer-based encoder network can learn both linguistic and acoustic information because the RNN-T is trained in an end-to-end manner. Therefore, restricting the Conformer to using only the local information can limit the Conformer’s potential modeling ability of the linguistic information inherent in the speech signal because some global information is important for linguistic modeling \cite{Rewon-GLS19}.

\begin{figure}[t]
  \centering
  \includegraphics[width=6.2cm,keepaspectratio]{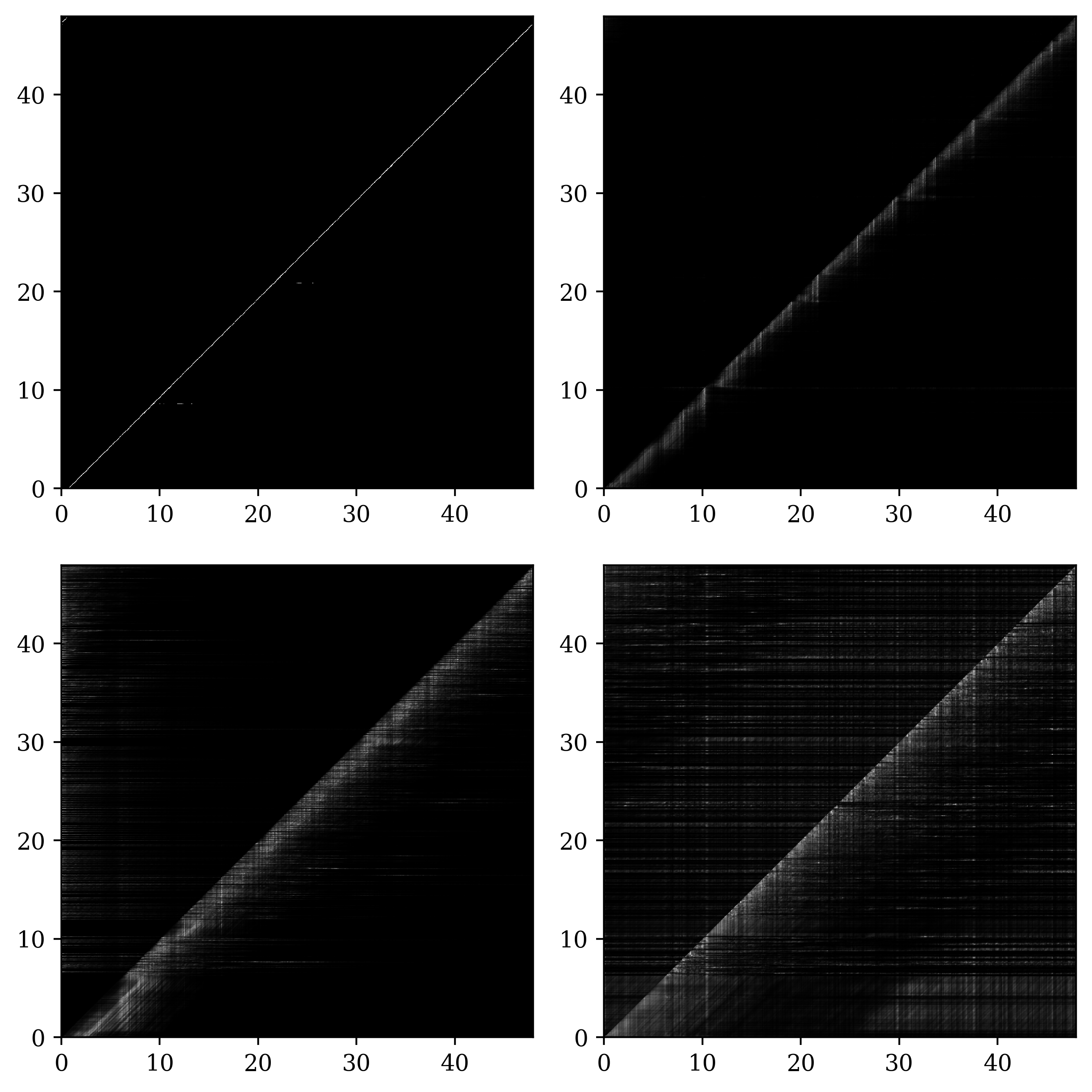}
  \caption{Explorations of attention behaviors for four self-attention layers: $1^{st}$ (top left), $6^{th}$ (top right), $10^{th}$ (bottom left), $12^{th}$ (bottom right). The x- and y-axes correspond to the the keys and queries, respectively. The numbers near the axes represent the time in seconds. This experiment is conducted with DOI-48, respectively.}
  \label{figure2}
\end{figure}

To validate the Conformer's utilization way for global information, we describe the attention behaviors of the trained Conformer as shown in Figure~\ref{figure2}. Considering Figure~\ref{figure2}, all the four layers pay high attention to the local information, whereas the higher layers ($10^{th}$ and $12^{th}$) attempt to deploy some global information together with sparse attention pattern. 

Based on this investigation, we decide to prune the attention scores while conserving the local and high-scored global attention in the inference phase through LM and sparse global mask (SGM) of which the details will be described in the next section. 
We believe this pruning improves the generalization ability because the attention scores of self-attention layers are the results of some computations with parameters related to query, key and value; thus, injecting the sparsity to the attention scores will reduce the involved parameters for the inference. Note that reducing the number of parameters is the classic approach to improve the generalization ability \cite{Giles-PRNN94,bishop-PRML06}. In Figure~\ref{figure1}, we also investigated the effectiveness of LM+SGM per the DOI length, showing consistent CER and deletion error improvement regardless of DOI length, which implies that using both local and some important sparse global connections is more appropriate than using only local connections or full connections to leverage the Conformer's potential local-global context modeling ability while maintaining the generalization ability. 

Our assumption is based on the investigation in Figure~\ref{figure2}. Therefore, we apply the pruning to the self-attention layers only to the inference phase, whereas the training is performed with full connections because if we apply pruning to the training phase, the observed attention patterns in Figure~\ref{figure2} cannot be guaranteed, which has also been confirmed by \cite{Rewon-GLS19}. 

\subsubsection{Pruning method for the sparsity}

%Based on this investigation, we assume that using both local and some important sparse global connections is more appropriate than using only local connections or a full connection to leverage the Conformer's potential local-global context modeling ability while maintaining the generalization ability. Our assumption is based on the investigation in Fig. 1. Therefore, we apply the proposed sparse self-attention layer only to the inference phase, whereas the training is performed with full connections because if we apply the sparse self-attention layer to the training phase, the observed attention patterns in Fig. 1 cannot be guaranteed, which has also been confirmed by [23].

Given an input sequence $\textbf{z}=\left[z_{1}, \cdots, z_{T}\right]$, the proposed sparse self-attention layer is constructed as follows:
\begin{equation}
\text{Attend}\left(z_{i}, S_{i}\right)=\text{softmax}\left(e_{i, j}\right)\cdot V_{j}, \text{where }{j\in S_i}
\end{equation}
\begin{equation}
e_{i,j}={\frac{\left(W_qz_i\right)\cdot K^T_j}{\sqrt{d}}}
\end{equation}
\begin{equation}
K_{j}=\left(W_k z_j\right), V_{j}=\left(W_v z_j\right)
\end{equation}
where $S_i$ is the attention mask, the set of indices of input vectors to be attended by $z_i$; $W_q$, $W_k$, and $W_v$ are the weight matrices that transform a given $z_i$ into a query, key, and value; $d$ is the inner dimension of the queries and keys. To leverage both the local and global connections, $S_i$ is designed as:
\begin{equation}
S_i=L_i\cup G_i
\end{equation}
where $L_i$ and $G_i$ are local and global masks, respectively. These are defined as $L_i=\left\{ j: i-w\le j \le i+w \right\}$ and $G_i=\left\{ j: \mu_i < e_{i,j} \right\}$; $w$ is a hyperparameter, and $\mu_i$ is the averaged attention score which is calculated as:
\begin{equation}
\mu_i=\frac{1}{T}\sum_{j=1}^T e_{i,j}
\end{equation}

 In the case of multi-head attention, we consider three types of global masks with the same local mask ($L_i$) for the $h$-head attention mask $S^h_i$ to investigate the generalization ability according to the degree of sparsity: $G^h_i$, $G^{and}_i=\bigcap^H_{h=1}G^h_i$, and $G^{or}_i=\bigcup^H_{h=1}G^h_i$, where $H$ is the number of attention heads. Regarding the three global attention types, $G^{and}_i$ will have the highest sparsity, whereas $G^{or}_i$ have the lowest. The final attention was performed as follows:
 \begin{equation}
 \text{attention}(\textbf{z})=W_p\left(\text{Attend}\left(z_i, S^h_i\right)\right)_{h\in\left\{1,\cdots,H\right\}}
 \end{equation}
where $W_p$ denotes the post-attention weight matrix.

\subsection{State reset at the silence for prediction network}

\begin{figure}[t]
  \centering
  \includegraphics[width=\linewidth]{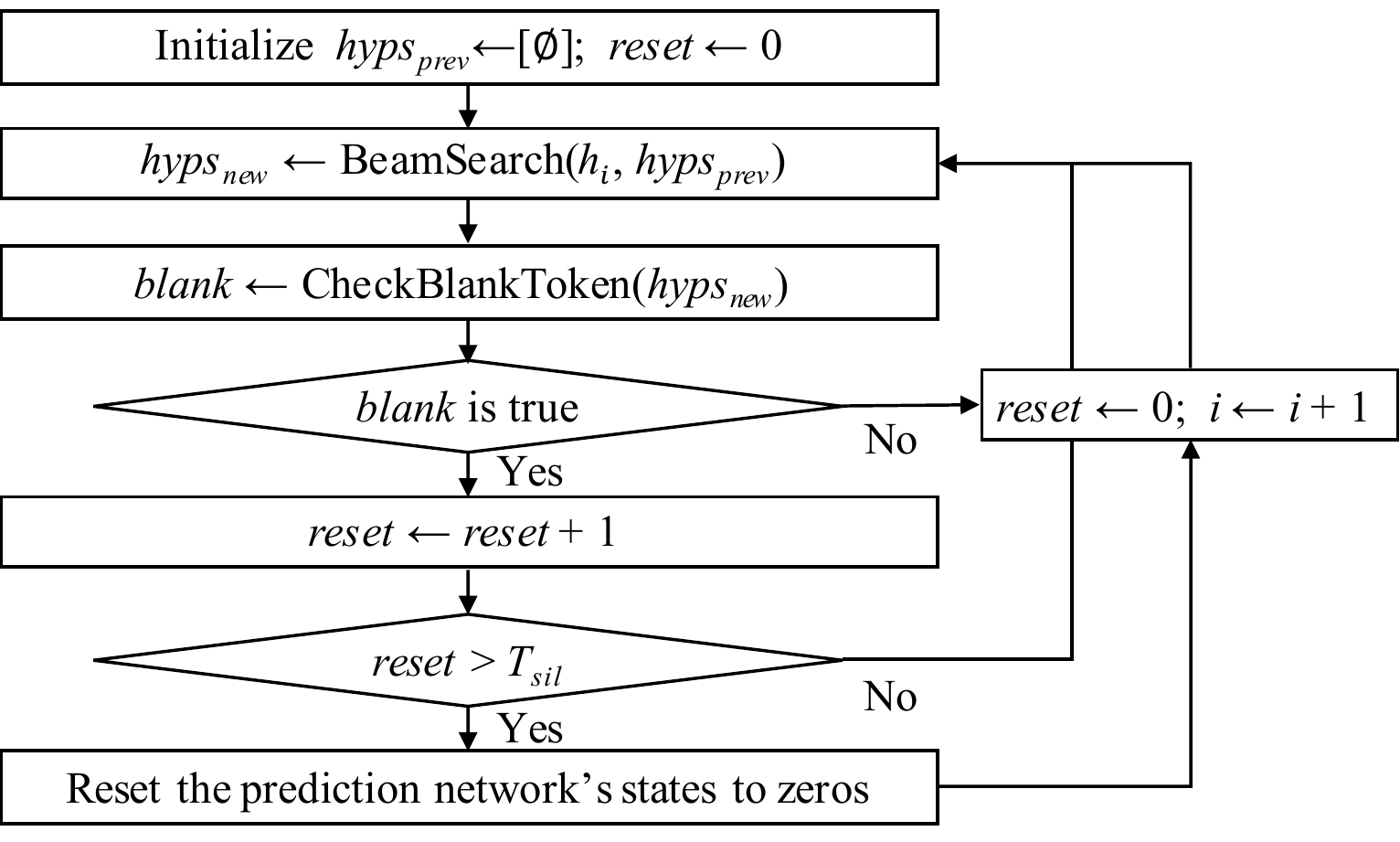}
  \caption{SRS block diagram.}
  \label{figure3}
\end{figure}

The prediction network can also be vulnerable to out-domain long-form utterances in the inference phase as the unseen linguistic context can be excessively accumulated in the prediction network. To alleviate this problem, it can be useful to reset the prediction network's states during the inference phase at the silent audio inputs. Thus, we propose a state reset method at the silence (SRS) as shown in Figure~\ref{figure3}. Given the encoder network outputs $\textbf{h}=\left[h_1, \cdots, h_T\right]$, $h_i$ is used for the beam search with the previous hypotheses $hyps_{prev}$. CheckBlankToken outputs `true' if all the last tokens from hypotheses in $hyps_{new}$ correspond to the blank tokens. If \textit{blank} is `false', \textit{reset} is set to zero and \textit{i} is increased by one for the next beam search. When it is `true', \textit{reset} is increased by one. If \textit{reset} is higher than the predefined hyperparameter $\textit{T}_{sil}$, the states of the prediction network are reset to zeros. We define consecutive blank tokens during the beam search as silence; thus, $\textit{T}_{sil}$ is related to the silence length. The aforementioned procedures are repeated until \textit{i} is equal to \textit{T}.

In Figure~\ref{figure1}, we also investigated the effectiveness of SRS, showing additional improvement compared to LM+SGM in CER and deletion error, though SRS shows some degradation in substitution error. This implies that prediction network also has a moderate in-domain overfitting issue.

\subsection{Segmentation methods for long-form utterances}

Although we adopt DOI to deal with long-form utterances, the DOI can degrade the long-term context if it splits the middle of the utterance even though there is an overlapped region between successive segments. As our sparse self-attention layer exploits some global information, we consider conserving the long-term context. Consequently, we also investigate an end-point detection (EPD) as another segmentation method. This is more likely to conserve the intact utterance, whereas the utterance length cannot be guaranteed.

%The out-domain long-form utterances not in the range of the training phase can cause some deletion errors during the inference phase \cite{Chiu-RMF21}. We were also suffered from the deletion problem with the Conformer RNN-T. To address this problem, we adopted two types of segmentation methods. The first is the DOI, which splits a long-form utterance into several uniform length utterances. Further, they partially overlap to alleviate the discontinuity problem between the segmented utterances. The DOI guarantees the utterance length; thus, input utterances are easily generalized by the RNN-T, considering the length. The DOI can degrade the long-term context if it splits the middle of the utterance although there is an overlapped region between the segmented utterances. As our self-attention layer exploits some global information, we consider conserving the long-term context. Consequently, we also adopt an end-point detection (EPD) as another segment method. This is more likely to conserve the intact utterance, whereas the utterance length cannot be guaranteed. This indicates that the utterance length detected by the EPD can be longer than our desired length.

\section{Experiments and Results}

\subsection{Experimental setup}

Our in-house dataset consists of 15k h of 10M Korean speech utterances. Most of them are related to the voice search domain for our voice assistant service. The utterances are generally short: the $50^{th}$ and $90^{th}$ percentile lengths are 2.5 s and 5.9 s, respectively. From the in-house dataset, approximately 1 h and 24 h of mutually exclusive utterances were randomly extracted for the validation and in-domain test sets, respectively. The remaining utterances were used for the training. Additional noise was mixed into the training set to make the overall signal-to-noise ratio (SNR) be between 5 dB and 20 dB. The length of the utterances in the in-domain test set ranged from 1 s to 10 s. For the out-domain test set, we collected approximately 24 h of videos from the broadcast (e.g., news, documentary, and variety show), and each video length ranged from 3 min to 50 min. All the datasets were anonymized and hand-transcribed.

The model implementation was based on the ESPNet toolkit \cite{Shinji-ESPnet18}. As input features, we used globally normalized 80- and 3-dimensional log Mel-filter bank coefficients and pitch features (83 dimensions in total), computed with a 25 ms window, shifted every 10 ms. The input features were first processed by a convolution subsampling layer, i.e., a 2-layer convolutional neural network with 256 channels, stride of 2, and a kernel size of 3, before forwarding them to the encoder network. The Conformer-based encoder network consists of 12 self-attention layers, and each layer has 1024 hidden units. We used 1-layer LSTMs with 640 cells as the prediction network and a joint network with 640 hidden units. As output labels, 256-word pieces based on Jamo (Korean alphabet) were used. The other model specifications and training strategy can be found in \cite{Guo-RDE21}. 

Considering the segmentation methods, the DOI and voice activity detection \cite{Kim-VAD18}-based EPD were used. After applying the EPD, the utterance lengths of in- and out-domain test sets were in the range of 0.4--2.48 s and 0.6--196.8 s, respectively. During the inference phase, we used a beam search described in \cite{Kim-ART20}. The hyperparameters $w$ and $\textit{T}_{sil}$ were set to 40 and 15, respectively, as determined from our validation set. The CER was used as an evaluation metric.

\subsection{Experimental results and discussion}

Table~\ref{table1} compares three types of sparse global masks: $G^{or}_i$ ($\text{SGM}_{1}$), $G^{h}_i$ ($\text{SGM}_{2}$), and $G^{and}_i$ ($\text{SGM}_{3}$) to the LM. The baseline utilizes full connections, without masks. $\text{SGM}_{3}$ has the highest sparsity among the three types. Considering the in-domain test set, the baseline showed the lowest CER, while exhibiting the highest CER in the out-domain test set. Applying LM to the baseline highly improved the CER in the out-domain test set, while degrading the CER in the in-domain test set. This implies that most of the global connections in the self-attention layers were fitted to the training domain, while local connections were relatively generalized to other domains. The addition of $\text{SGM}_{1}$ and $\text{SGM}_{2}$ to the LM resulted in a lower CER in the out-domain test set than the baseline, while keeping the CER similar to the baseline in the in-domain test set. However, both $\text{LM}\text{+SGM}_{1}$ and $\text{LM}\text{+SGM}_{2}$ showed higher CER than the LM in the out-domain test set, implying that $\text{SGM}_{1}$ and $\text{SGM}_{2}$ still have some global connections overfit to the training domain. Considering the comparison between $\text{LM}\text{+SGM}_{3}$ and LM, we can claim that some global context information captured by $\text{SGM}_{3}$ is valid in other domains. This indicates that the global connections concurrently showing high attention scores in all attention heads are generalized because using these connections improves the CER in both the in- and out-domain test sets. We refer to $\text{SGM}_{3}$ as the SGM for the rest of this study.

\begin{table}[t]
  \caption{Comparison of CER with different global sparsity patterns on in- and out-domain test sets}
  \label{table1}
  \centering
  \begin{tabular}{ l c c }
    \toprule
    \text{Model} & \text{In-Domain} & \text{Out-Domain} \\
    \midrule
    $\text{Baseline}$ & $\textbf{6.25}$ & $17.16$~~~ \\
    $\text{LM}$ & $7.16$  & $13.84$~~~ \\
    + $\text{SGM}_{1}$ & $6.26$  & $16.57$~~~ \\
    + $\text{SGM}_{2}$ & $6.26$  & $16.70$~~~ \\
    + $\text{SGM}_{3}$ & $6.92$  & $\textbf{12.99}$~~~              \\
    \bottomrule
    \multicolumn{3}{l}{The numbers in bold indicate the best result.} \\
    \multicolumn{3}{l}{EPD was used for the segmentation method.}\\
  \end{tabular}
  
\end{table}

\begin{table}[t]
  \caption{Comparison of CER on in- and out-domain test sets}
  \label{table2}
  \centering
  \begin{tabular}{l l l l l }
    \toprule
    \multirow{2}{*}{Model} & \multicolumn{2}{c}{In-Domain} & \multicolumn{2}{c}{Out-Domain} \\ \cline{2-5}
    & \multicolumn{1}{c}{DOI} & \multicolumn{1}{c}{EPD} & \multicolumn{1}{c}{DOI} & \multicolumn{1}{c}{EPD} \\
    \midrule
    $\text{Baseline}$ & $\textbf{6.11}$ & $\textbf{6.25}$ & $15.72$ & $17.16$~~~ \\
    + $\text{SRS}$ & $6.39$ & $6.57$ & $15.05$ & $16.73$~~~ \\
    $\text{LM}$ & $6.83$ & $7.16$ & $13.50$ & $13.84$ ~~~ \\
    + $\text{SGM}$ & $6.76$ & $6.92$ & $13.33$ & $12.99$~~~ \\
    %+ SRS & 7.00 & 7.32 & 13.51 & 13.63~~~ \\
    + $\text{SGM}$ + $\text{SRS}$ & $6.92$ & $6.97$ & $\textbf{13.27}$ & $\textbf{12.94}$~~~ \\
 
    \bottomrule
    \multicolumn{5}{l}{The numbers in bold indicate the best result.}\\
  \end{tabular}
  
\end{table}

Table~\ref{table2} compares the proposed methods (SGM and SRS) to the baseline and LM based on the segmentation methods (DOI and EPD). For DOI, DOI-20 was used. When the SGM was applied to the LM, the CER was lower than the LM for all the segmentation methods and test sets. To leverage the global context information, it is desirable to use an intact utterance without splitting. For the in-domain test set, all the utterances were not affected by the DOI as all of them were shorter than 20 s, whereas most of the utterances in the out-domain test set were split by the DOI. Thus, the SGM with EPD showed lower CER in out-domain utterances because the EPD split the out-domain utterances less than the DOI. 

Applying the SRS improved performance on all the cases in the out-domain test set; thus, LM+SGM+SRS with EPD achieved 24.6\% and 6.5\% relative CER reduction compared to the baseline and LM, respectively, while degrading the CER in the in-domain test set. This implies that the long-term linguistic context modeling by prediction network has an overfitting issue to the in-domain; thus, splitting the long-term context modeling to the short-term by SRS can be helpful for generalization.

In Table~\ref{table2}, we observe that LM+SGM+SRS with DOI achieves only a 1.7\% relative CER reduction compared to the LM. We assume that the DOI-20 is too short to leverage the global context information by the SGM. To validate this assumption, we perform the experiments according to various DOI lengths: 20, 28, 38, and 48 as shown in Table~\ref{table3}. In this experiment, the 2 s overlap was maintained, and we only increased the segment length. The DOI-48 is the maximum length as our self-attention layers only consider the past and future 24 s as described in Figure~\ref{figure2}. In Table~\ref{table3}, LM+SGM+SRS shows further CER reduction as the DOI length increases; LM+SGM+SRS with DOI-48 achieves 27.6\% and 5.3\% relative CER reduction compared to the baseline and LM, respectively. This result implies that the effectiveness of the SGM and SRS is exhibited when the utterance length is sufficiently long.

\begin{table}[t]
  \caption{Comparison of CER on the out-domain test sets based on the DOI length (s)}
  \label{table3}
  \centering
  \begin{tabular}{ l l l l l }
    \toprule
    \multirow{2}{*}{Model} & \multicolumn{4}{c}{DOI Length} \\ \cline{2-5} 
    & \multicolumn{1}{c}{20} & \multicolumn{1}{c}{28} & \multicolumn{1}{c}{38} & \multicolumn{1}{c}{48} \\
    \midrule
    $\text{Baseline}$ & $\textbf{15.72}$ & $15.75$ & $16.95$ & $17.71$~~~ \\
    $\text{LM}$ & $\textbf{13.50}$ & $13.68$ & $13.63$ & $13.54$ ~~~ \\
    + $\text{SGM}$ & $13.33$ & $13.48$ & $\textbf{13.04}$ & $13.06$~~~ \\
    %+ SRS & 7.00 & 7.32 & 13.51 & 13.63~~~ \\
    + $\text{SGM}$ + $\text{SRS}$ & $13.27$ & $13.38$ & $12.97$ & $\textbf{12.82}$~~~ \\
    \bottomrule
    \multicolumn{5}{l}{The numbers in bold indicate the best result.}\\
  \end{tabular}
  
\end{table}

\begin{figure}[t]
  \centering
  \includegraphics[width=7cm,keepaspectratio]{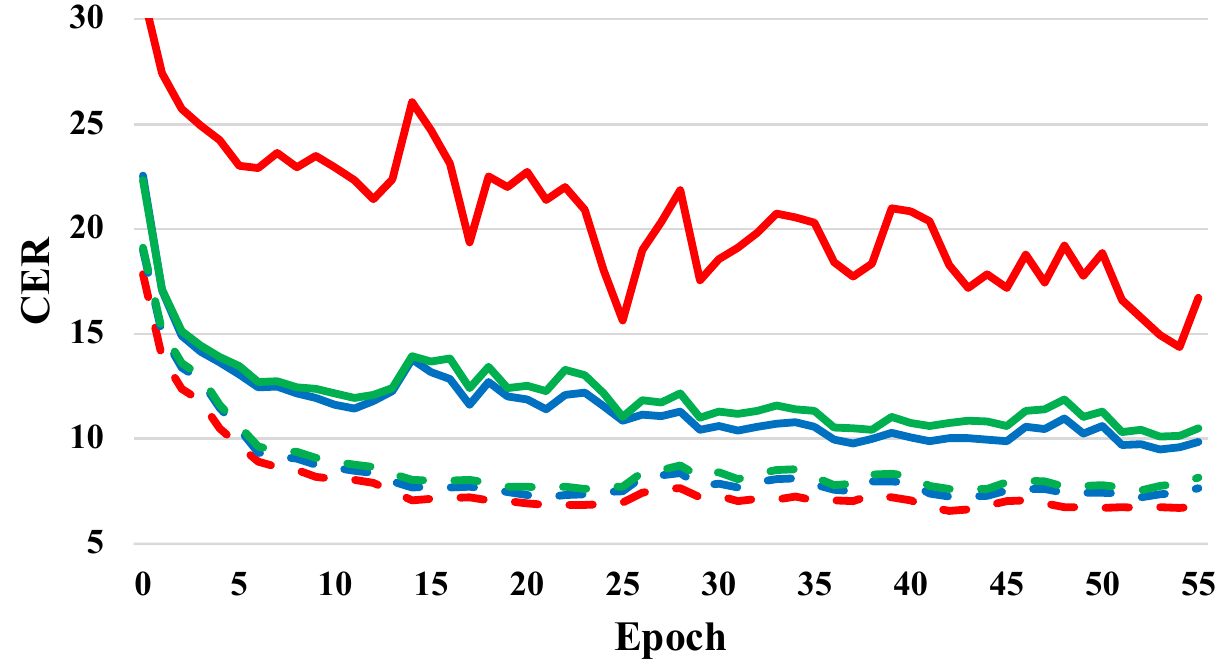}
  \caption{CERs considering the mask types: no mask (red), LM (green), and LM+SGM (blue) on the in- (dashed line) and out- (solid line) domain test sets as a function of training epochs. EPD was used for the segmentation method.}
  \label{figure4}
\end{figure}

In Figure~\ref{figure4}, we investigate the effect of the LM and LM+SGM based on the training epochs. All the masks were applied only during the inference phase. The baseline (no mask) showed the lowest CER across all the training procedures in the in-domain test set and the highest CER in the out-domain test set. Further, the baseline exhibited high CER variation in the out-domain test set consistently, while showing a low variation in the in-domain test set. This indicates that the baseline suffers from the generalization problem across domains. In contrast, both the LM and LM+SGM showed stable CER decreases during the training in both the in- and out-domain test sets. Considering the earlier training epochs ($<5$), there was no obvious CER difference between the LM and LM+SGM. Regarding the latter training epochs ($>25$), however, we found that the CER difference between the LM and LM+SGM was consistently maintained in both the in- and out-domain test sets, implying that they learn the local context information first and learn the global one later.

%\textbf{Librispeech}: We also conducted the experiments for RNN-T trained on Librispeech. We just followed the experimental setup implemented in [28]. For the test sets, we used Librispeech’s test sets (in-domain) and approximately 10 hours of videos collected from YouTube (out-domain); each video length was in the range from 1 to 50 minutes. As the results of our in-house Korean dataset, the baseline showed the best WER in in-domain test set while the worst in out-domain test set. The LM improved the out-domain WER, however, showed degradation of WER in in-domain. Applying SGM to LM consistently improved WER in all test sets. The addition of SRS showed lower WER in out-domain test set while degrading WER in in-domain.

\section{Conclusion}
We proposed the sparse self-attention layers and a state reset method at the silence (SRS) to alleviate the in-domain overfitting issue for Conformer RNN-T’s encoder and prediction networks, respectively. We found that most of connections in self-attention layers overfit to the in-domain, causing high deletion errors due to the loss of generalization ability. This motivates us to design local and sparse global masks (LM+SGM) to prune the most of overfit connections while remaining local and high-scored global connections considered to be generalized. Further, we prune the excessive contextual information for prediction network with SRS as learned long-term contextual modeling of prediction network can be closed to the in-domain. When the DOI length is 48, the Conformer RNN-T equipped with LM+SGM and SRS outperformed the vanilla Conformer RNN-T with a relative CER of 27.6\% in out-domain test sets.

%We proposed sparse self-attention layers to generalize RNN-T inference by using local and global attention masks. The local and three types of sparse global self-attention masks were compared for Conformer-based encoder networks and obtained low CER comparing to the full connections. Our results demonstrate that injecting appropriate sparsity on the connections leads to better performance in domain mismatch situations. We obtained further improvement in long utterances from state reset method and segment methods by pruning the accumulated context that can be redundant. In this study, we focused on the scores of each attention head individually; future works will explore alternative generalization techniques that address the relationship between attention heads.

\bibliographystyle{IEEEtran}

\bibliography{sparseattn}

\end{document}